\documentclass[oldversion,rnote]{aa} % double column
\usepackage{graphicx}
\usepackage{longtable}
\usepackage{txfonts}
\usepackage{rotating}
\usepackage{lscape}
\newcommand{\gsim}{\;\lower.6ex\hbox{$\sim$}\kern-7.75pt\raise.65ex\hbox{$>$}\;}
\newcommand{\lsim}{\;\lower.6ex\hbox{$\sim$}\kern-7.75pt\raise.65ex\hbox{$<$}\;}

\begin{document}
\title{The connection between missing AGB stars and extended horizontal branches
%\thanks{Based on observations collected at 
%ESO telescopes under programme 073.D-211}
 }

\author{
R.G. Gratton\inst{1},
V. D'Orazi\inst{1},
A. Bragaglia\inst{2},
E. Carretta\inst{2},
\and
S. Lucatello\inst{1,3}
%M. Bellazzini\inst{1},
%G. Catanzaro\inst{4},
%F. Leone\inst{5},
%Y. Momany\inst{2,6},
%G. Piotto\inst{7}
%\and
}

\authorrunning{R.G. Gratton}
\titlerunning{Missing AGB stars and extended HBs}

\offprints{R.G. Gratton, raffaele.gratton@oapd.inaf.it}

\institute{
INAF-Osservatorio Astronomico di Padova, Vicolo dell'Osservatorio 5, I-35122
 Padova, Italy
\and
INAF-Osservatorio Astronomico di Bologna, Via Ranzani 1, I-40127
 Bologna, Italy
\and
Excellence Cluster Universe, Technische Universit\"at M\"unchen, 
Boltzmannstr. 2, D-85748, Garching, Germany 
  }

\date{}

\abstract{Recent surveys confirm early results about a deficiency or
even absence of CN-strong stars on the asymptotic giant branch (AGB) of 
globular clusters (GCs), although with quite large cluster-to-cluster 
variations. In general, this is at odds with the distribution 
of CN band strengths among first ascent red giant branch (RGB) stars.
Norris et al. proposed that the lack of CN-strong stars in
some clusters is a consequence of a smaller mass of these stars
that cannot evolve through the full AGB phase. In this short paper we  
found that the relative frequency of AGB stars can change by a factor of two 
between different clusters. We also find a very good correlation between 
the minimum mass of stars along the horizontal branch (Gratton et al. 
2010) and the relative frequency of AGB stars, with a further dependence
on metallicity. We conclude that indeed
the stars with the smallest mass on the HB cannot evolve through the full
AGB phase, being AGB-manqu\'e. These stars likely had large He and N 
content, and large O-depletion. We then argue that there should not be 
AGB stars with extreme O depletion, and few of them with a moderate one.  }
\keywords{Stars: abundances -- Stars: evolution --
Stars: Population II -- Galaxy: globular clusters }

\maketitle

\section{Introduction}

Stars with very strong CN bands (CN-strong stars) are very common along the 
red giant branch (RGB) of globular clusters (GCs). It is then quite curious that
most asymptotic giant branch (AGB) stars in GCs have weak CN bands, a fact first 
discovered by Norris et al. (1981) in the case of NGC~6752 (see also the 
review by Sneden et al. 2000). Similar results were obtained from Na and Mg abundances for a large
sample of giants in M13 by Pilachowski et al. (1996): stars with high
[Na/Fe], dominating the upper RGB, do not seem to be represented among
the AGB sample. Very recently, 
Campbell et al. (2010) reported initial results of a medium resolution survey 
of about 250 AGB stars over 9 GCs. These early results confirmed the lack of 
CN-strong AGB stars in NGC~6752, while in other GCs (like M5: Smith \& Norris 
1993; and possibly like 47~Tuc: Mallia 1978, Campbell et al. 2006) a few 
CN-strong AGB stars are present, although they are less frequent 
than among RGB stars.

The low incidence of CN-strong stars along the AGB is striking, since
the presence of strong CN bands is usually assumed to be evidence of
more mixed material. However, a very interesting interpretation of this
phenomenon was proposed by Norris et al. (1981). According to this 
hypothesis, "when star formation ceased in the cluster, there were two groups
of stars having not only the observed carbon and nitrogen properties, but
also a difference in helium abundance, $\Delta Y\sim 0.05$, in the sense
that the nitrogen strong group has enhanced helium. This difference in 
helium leads to a mass difference of $\sim 0.07~M_\odot$\ at the main
sequence turn-off, which, together with our current knowledge of
horizontal branch morphology, provides an explanation of both the gap
on the horizontal branch and the lack of CN-strong stars on the AGB. (The high helium, 
high CN group does not ascend the giant branch for a second time)``.
This possible explanation was supported by models of HB stars
(see e.g. Sweigart \& Gross 1976; Greggio \& Renzini, 1990; Lee et al. 1994).

Thirty years later, we now know that multiple generations, differing in
Na, O, and likely also He abundances, exist in virtually all GCs (Carretta et 
al. 2009a). In a separate paper (Gratton et al. 2010) we discussed the
connection of the multiple generation scenarios with the morphology of
the HB. Following the arguments made by Ventura et
al. (2001), we interpret the extension of the horizontal branch (HB) to low masses observed 
in several GCs as the effect of enhanced He abundances (which combine
with differences in ages and possibly other parameters to explain the
so-called second parameter issue), exactly as proposed by Norris et al. 
(although with a different scenario for the formation of GCs: see 
Carretta et al. 2010). 

Within this framework, we should then expect that there is a close
relation between the extension of the HB (as represented e.g. by the
minimum mass along the HB: see Gratton et al. 2010) and the frequency 
of AGB stars (as given by the ratio between the number of stars along
the AGB and those on the RGB: see discussion in Sneden et al. 2000).
While M13 indeed seems to have a low frequency of AGB stars (Caputo
et al. 1978; Buzzoni et al. 1983), a comprehensive picture is missing.
In this research note, we evaluate this correlation from current data 
(Sect. 2), and briefly discuss its implications (Sect. 3). Conclusions 
are drawn in Sect. 4.

\begin{center}
\begin{figure}
\includegraphics[width=8.8cm]{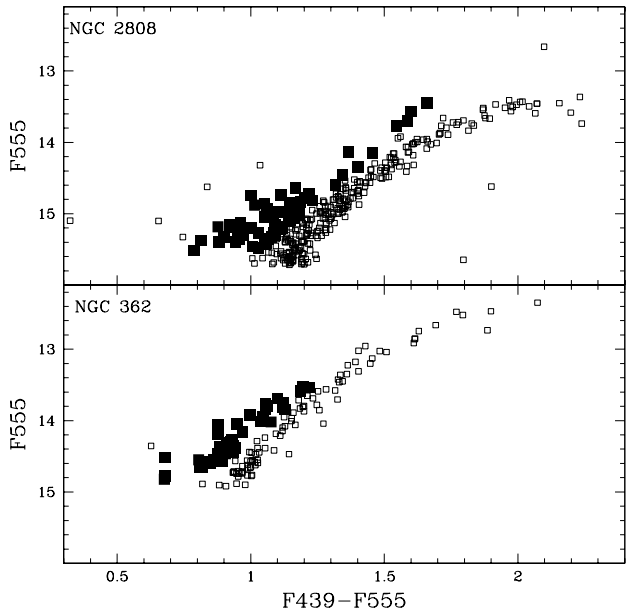}
\caption{Separation of AGB stars (filled squares) and other stars (small
empty squares) in the HST colour magnitude diagrams of NGC~2808 (upper 
panel) and NGC~362 (lower panel)}
\label{f:fig1}
\end{figure}
\end{center}

\begin{center}
\begin{figure}
\includegraphics[width=8.8cm]{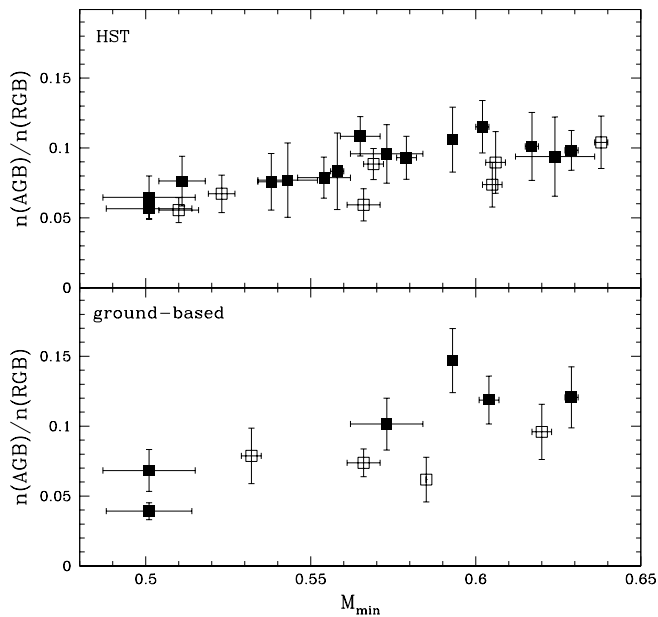}
\caption{Run of the $f_{\rm AGB}=n$(AGB)/$n$(RGB) ratio with the minimum mass along
the HB $M_{\rm min}$\ (see Gratton et al. 2010) for both HST (upper panel) 
and ground-based data (lower panel). Filled squares are for GCs with
[Fe/H]$>-1.7$, open squares for GCs with [Fe/H]$<-1.7$. }
\label{f:fig2}
\end{figure}
\end{center}

\section{The correlation between the minimum mass on the HB and the frequency of AGB stars}

We evaluated the frequency of AGB stars in a number of GCs from the
high quality photometric data sets considered by Gratton et al. (2010).
These consist of two databases: the results of the HST snapshot survey by 
Piotto et al. (2002), and the ground-based survey of GCs by Rosenberg
et al. (2000a, 2000b). With respect to the analysis of HB stars presented in
Gratton et al. (2010) we have two additional problems. 

First, AGB stars are an order of magnitude less numerous than RGB and HB stars. Hence, 
only those GCs with large populations of RGB stars have a numerous enough
population of AGB stars. In practice, for HST data we limited ourselves only to
those GCs that have $>100$~RGB stars, where the definition of RGB stars
is here those stars more luminous than $V(HB)+1$\ (see Gratton et al. 2010).
For the ground-based data (where contamination by field stars is
larger) we used an even more restrictive criterion
of $>200$~RGB stars. The values of the magnitude of the HB $V(HB)$\ were taken
from Harris (1996).

Second, separation of AGB stars from RGB stars is very difficult in those
clusters where there is significant differential reddening. This is usually
the case for highly reddened clusters. Hence, we limited ourselves to
those GCs having a reddening $E(B-V)\leq 0.25$. 

The total of GCs satisfying these two criteria is 26, covering a wide range
in metallicity and extension of the HB. Low luminosity clusters are however
absent from our sample, simply because they do not have enough stars.

In each GC, we first subtracted the field stars, following the same procedure
described in Gratton et al. (2010). Then we separated AGB stars from those on 
the HB assuming that only stars with $V<V(HB)-0.5$\ could be on the AGB, and
then separated AGB stars from RGB stars by their colours being bluer than
a dividing straight line, whose inclination is a function of metallicity,
and horizontal position was guided by eye. Stars much bluer 
(more than 0.3 mag) than this dividing
line were assumed to be field object, not properly subtracted by our
procedure. We acknowledge that this procedure neglects a few bright AGB stars - 
which are anyway very difficult to be separated from the brightest RGB
stars. However, there are very few such stars in each GCs, so that the
impact of this error in our discussion is very small (actually, much smaller 
than our error bars). On the other hand, uncertainties in the separation of AGB and RGB
stars contribute to the errors. In the case of HST photometry, which is more
accurate, on average we change our estimates of the ratio between the number of 
AGB and RGB stars by $\pm 0.011$\ if we shift the colour of the separating line 
by $\pm 0.02$~mag. Since ground-based data are less accurate, we expect larger 
errors. Figure~\ref{f:fig1} illustrates a couple of examples of application of 
this procedure.

Table~\ref{t:tab1} gives the list of the GCs, their metallicity from Carretta et 
al. (2009b); the absolute magnitude $M_V$\ and the reddening E(B-V) from Harris 
(1996); the minimum mass along the HB $M_{\rm min}$\ and the number of RGB stars 
$n$(RGB) from Gratton et al. (2010); the number of AGB stars $n$(AGB), and the ratio $f_{\rm AGB}$
between the number of AGB and RGB stars, with the error given by Poisson statistics.
For NGC~2419, we adopted a value of $M_{\rm min}$=0.51~$M_\odot$, which is consistent
with its extreme BHB (Ripepi et al. 2007), below the limiting magnitude of the Piotto 
et al. (2002) photometry.

Figure~\ref{f:fig2} displays the run of $f_{\rm AGB}$\ with $M_{\rm min}$\ for both 
HST and ground-based data. As expected, in both cases there is a clear correlation 
between these two quantities. Considering only the HST data, the linear
correlation coefficient is $r=0.75$\ (22 GCs), which is significant at a very
high level of confidence. GCs with extended blue HBs (like NGC~2808) have 
roughly half the relative frequency of AGB stars of those with short HBs. This
confirms the earliest finding for M13 by Caputo et al. (1978) and Buzzoni et al.
(1983), which is now shown to be a general property. The correlation
seems to break at large values of $M_{\rm min}$. This might be due to a combination
of two effects: first, in our approach we neglected the most luminous AGB stars,
a phase that can be reached only by the evolution of the most massive HB stars in GCs;
and second, the lifetime on the AGB of the most massive stars in GCs might
be only weakly dependent on the mass of the star on the HB, perhaps because the
mass loss rate is very large for the most luminous stars.

As noticed by the referee, while the correlation is indeed good, there seems
to be a significant spread in $f_{\rm AGB}$\ values at a given $M_{\rm min}$.
This spread is correlated with cluster metallicity (see Figure~\ref{f:fig2}): in fact, the correlation
between [Fe/H] values and residuals around the best fit line is $r=0.52$\ (22 GCs), 
which is significant at better than 1\%.

The correlation between the frequency of AGB stars and metallicity should
not come as a surprise. In fact, Frogel \& Elias (1988) found that the
maximum luminosity of stars along the AGB is a function of metallicity
(see their Fig. 1). This agrees with a scenario where the evolution of
small mass stars along the AGB is limited by the first thermal pulse,
which occurs at lower luminosities (and masses) for metal-poor stars (Renzini
\& Fusi Pecci, 1988). However, the correlation with metallicity is fairly
independent from that on $M_{\rm min}$, which is stronger. This correlation
merits then a separate discussion.

\section{Discussion}

We first notice that the ratio $f_{\rm AGB}\sim 0.1$\ we obtain for 
those GCs rich in AGB stars agrees with expectations based on lifetimes of 
the corresponding evolutionary phases. These are 14 Myr for a 0.6~$M_\odot$\ 
AGB stars, and 140 Myr for the time required for a moderately metal-poor 
($Z=10^{-3}$) star of 0.8~$M_\odot$\ to climb the RGB from 1 magnitude below 
the HB level, up to the tip of the RGB (see Bertelli et al. 2009). On the
other hand, the obvious interpretation of the correlation between $n$(AGB)/$n$(RGB)
and $M_{\rm min}$\ is that small mass HB stars either do not reach the AGB, 
or leave it much earlier than the more massive ones.

To give further insight into this issue, we consider the following rough
argument. Using the core mass-luminosity relation (Paczynski 1971, Marigo 2000)
and stellar models (Bertelli 2009), we can estimate that the core mass of a 
GC star when it reaches the AGB is about 0.49~$M_\odot$, and it is about
0.50~$M_\odot$ when it reaches the luminosity similar to that of the tip of the RGB
(about $\log{L/L_\odot}\sim 3$), which is roughly the maximum observed 
luminosity for stars in GCs (actually, this luminosity depends
on metallicity; see previous Section). According to models, the mass of the envelope of 
a 0.6~$M_\odot$\ AGB stars with a luminosity of $\log{L/L_\odot}\sim 3$ is 
$\sim 0.05$~$M_\odot$. Let us then assume that a star leaves the AGB when the
mass of the envelope is that large\footnote{ The criterion considered by
Renzini \& Fusi Pecci (1988) is slightly different, being 0.04~$M_\odot$ the
mass between the two burning shells.}. Hence an 
HB star with a mass $<0.54$~$M_\odot$\ will not even start the AGB.
The same models indicate that a star of 0.8~M$_\odot$ and $Z=10^{-3}$\
takes about 29~Myr to climb the RGB from the luminosity of the base of the
AGB up to the tip of the RGB, while a 0.6~M$_\odot$\ AGB star takes about
14~Myr to run the corresponding track along the AGB. According to Gratton
et al. (2010), a typical RGB star loses some 0.2~$M_\odot$\ before reaching
the HB (see also Renzini \& Fusi Pecci 1988); most of 
this loss is in the late phases of the RGB. Assuming that
RGB and AGB stars have a similar mass loss rate, we may then expect that
a GC star may lose some 0.1~$M_\odot$\ while climbing the AGB. Then, the
minimum HB mass required to reach the tip of the AGB is about 0.65~M$_\odot$,
which is indeed the typical mass of HB stars (see Fig. 10 of Gratton et
al. 2010). (This is actually a sanity check, showing that our reasoning
is consistent with the data we have). According to this picture, small 
mass HB stars ($M<0.54~M_\odot$) do not begin the AGB. These objects
are the AGB-manqu\'e stars (Greggio \& Renzini 1990); they have been 
found in several clusters (see e.g. the case of NGC~2808: Castellani et al. 
2006). The others leave the AGB at different luminosities, and only those 
with masses $M>0.65~M_\odot$ may get as luminous as the tip of the RGB. We 
could then expect that the frequency of AGB stars depends on the distribution
of masses along the HB.

We note that if the mass of an RGB star is smaller than that required for the 
He-core flash (about 0.50~$M_\odot$: Castellani \& Castellani 1993), the star 
will leave the RGB before reaching its tip. Such stars, usually called 
RGB-manqu\'e, will become He-white dwarfs; however, they will likely have a 
late He-flash, after which they will move to the blue hook of the HB, and 
later once for all to the C-O white dwarf cooling sequence, without becoming an AGB
star (Castellani \& Castellani 1993, D'Cruz et al. 2000, Brown et al. 2001, 
Moehler et al. 2004). These small mass RGB stars might either be in mass transfer 
binaries, or normal single stars with a suitable combination of helium abundance, 
age, and metal abundance. For instance, using data and equations considered in 
Gratton et al. (2010), single stars in NGC~2808 (or in M~13) with roughly $Y>0.35$\ ($Y>0.335$)
should be RGB-manqu\'e, those with $0.33<Y<0.35$ ($0.31<Y<0.335$) should become AGB-manqu\'e,
while those with $Y<0.33$ ($Y<0.31$) might successfully start their evolution along the AGB.
Only those AGB stars with $Y\sim 0.25$\ will reach a luminosity similar to the tip 
of the RGB. However, most of the mass loss by RGB stars occurs just in the latest
phases of the RGB. Castellani \& Castellani (1993), considered the Reimers (1975) mass
loss law with different values of the efficiency parameter $\eta$. The values of $\eta$\ 
they considered were quite large. The smallest ones, with $\eta=0.5$\ yields an
average mass loss of 0.2~$M_\odot$\ along the RGB of a metal-poor GC like M15, somewhat
larger than required to explain its HB (Gratton et al. 2010). In such a model, half
of the mass is lost in the last 0.4 Myr, that is $<1$\% of the time spent at 
luminosities brighter than $M_V({\rm HB})+1$, while the star is climbing the last 
0.7 mag (in bolometric magnitude) on the RGB. Other mass loss laws give an even stronger 
dependence on luminosity (see the discussion in Catelan 2009), and hence concentration 
of the mass loss in the later phases of the RGB. As a consequence, even the most He-rich 
stars are expected to make most of the evolution along the RGB. While there should be a 
deficiency of He-rich stars very close to the tip of the RGB, it will be very difficult
to establish this effect due to the small numbers involved. As a consequence, we should
not expect any significant effect on the number counts on the RGB, while of course
there are large consequences on the later evolutionary phases.

We may better evaluate the impact of our result by considering two GCs
with similar metallicity, but very different HB morphology. A similar pair
may be provided by NGC~362 and NGC~2808. The first cluster has a quite short
HB, with a minimum mass of 0.60~M$_\odot$, and a median one of 0.68~M$_\odot$.
In such a cluster, most stars will climb up through the whole AGB, and should
then have a large value of $f_{\rm AGB}$. Indeed we get $f_{\rm AGB}=0.115\pm 0.019$,
among the highest we found. We then expect that the distribution of stars
along the Na-O and other anti-correlations to be similar for AGB and RGB
stars (a prediction that could be tested with appropriate observations). On
the other hand, in the case of NGC~2808 the distribution of stars along the
HB is trimodal: about 20\% of the stars are extreme BHB stars, with masses 
well below the minimum of 0.54~$M_\odot$ required to start evolution along 
the AGB; about 40\% of the stars are quite massive ($\sim 0.68~M_\odot$),
and similarly to the bulk of the stars of NGC~362 they may evolve through
the AGB up to a luminosity similar to that of the tip of the RGB; finally,
another 40\% of the stars have a mass of about 0.61~M$_\odot$, which should allow them
to start their evolution along the AGB, which is however terminated before
reaching the luminosity of the tip of the RGB. Assuming that these stars are able to be on the
AGB half the time of the more massive ones, we may expect that NGC~2808 has
a value of $f_{\rm AGB}$ which is about 60\% that of NGC~362. The
value we obtain is actually a bit lower, about $50\pm 11$\% using HST data, and even
lower using the ground-based data. This suggests that the progeny of HB stars
of intermediate mass of NGC~2808 are able to remain on the AGB for less than
half of the typical lifetime of AGB stars. In this case, we then expect that the distribution
of AGB stars along the various anti-correlations (C-N, Na-O, Mg-Al) be very
different from that found for RGB stars, because we expect that extreme BHB 
stars are connected with the extremely O-depleted stars, the BHB stars with the
moderately O-poor, and the RHB with the O-rich stars. If this picture is correct, 
we predict that there should be no extremely O-depleted stars, about a third or less of
moderately O-depleted, and a vast majority of O-rich stars along the AGB of NGC~2808.

\begin{table*}
\centering
\caption[]{Counts of AGB stars}
\begin{tabular}{cccccccc}
\hline
  NGC    &$[$Fe/H$]$ &$M_v$   & E(B-V) & $M_{\rm min}$ & $N_{\rm RGB}$ & $N_{\rm AGB}$ & $f_{\rm AGB}$   \\
\hline
\multicolumn{8}{c}{HST} \\
\hline
     104 & -0.76 & -9.42 &  0.04 & 0.629 &  529 &  52 & 0.098$\pm$0.014 \\ % 5 3
     362 & -1.16 & -8.41 &  0.03 & 0.602 &  365 &  42 & 0.115$\pm$0.019 \\ % 2 2
    1261 & -1.35 & -7.81 &  0.01 & 0.593 &  217 &  23 & 0.106$\pm$0.023 \\ % 1 1
    1851 & -1.22 & -8.33 &  0.02 & 0.579 &  430 &  40 & 0.093$\pm$0.015 \\ % 3 4
    1904 & -1.57 & -7.86 &  0.01 & 0.538 &  198 &  15 & 0.076$\pm$0.020 \\ % 0 1
    2419 & -2.12 & -9.58 &  0.11 & 0.510 &  739 &  41 & 0.056$\pm$0.009 \\ % 6 4
    2808 & -1.15 & -9.39 &  0.22 & 0.501 & 1042 &  59 & 0.057$\pm$0.008 \\ % 2 7
    5024 & -1.99 & -8.70 &  0.02 & 0.638 &  327 &  34 & 0.104$\pm$0.019 \\ % 6 2
    5634 & -1.88 & -7.69 &  0.05 & 0.606 &  201 &  18 & 0.090$\pm$0.022 \\ % 2 1
    5694 & -1.86 & -7.81 &  0.09 & 0.605 &  312 &  23 & 0.074$\pm$0.016 \\ % 0 0
    5824 & -1.85 & -8.84 &  0.13 & 0.569 &  780 &  69 & 0.088$\pm$0.011 \\ % 11 5
    5904 & -1.29 & -8.81 &  0.03 & 0.573 &  240 &  23 & 0.096$\pm$0.021 \\ % 0 1
    6093 & -1.75 & -8.23 &  0.18 & 0.523 &  402 &  27 & 0.067$\pm$0.013 \\ % 2 1
    6205 & -1.54 & -8.70 &  0.02 & 0.501 &  294 &  19 & 0.065$\pm$0.015 \\ % 2 1
    6229 & -1.43 & -8.05 &  0.01 & 0.554 &  394 &  31 & 0.079$\pm$0.015 \\ % 1 1
    6584 & -1.49 & -7.68 &  0.10 & 0.624 &  128 &  12 & 0.094$\pm$0.028 \\ % 1 0
    6637 & -0.70 & -7.64 &  0.16 & 0.617 &  188 &  19 & 0.101$\pm$0.024 \\ % 3 2
    6681 & -1.51 & -7.11 &  0.07 & 0.543 &  117 &   9 & 0.077$\pm$0.027 \\ % 0 0
    6723 & -1.12 & -7.84 &  0.05 & 0.558 &  120 &  10 & 0.083$\pm$0.027 \\ % 0 0
    6864 & -1.16 & -8.55 &  0.16 & 0.565 &  600 &  65 & 0.108$\pm$0.014 \\ % 2 3
    7078 & -2.26 & -9.17 &  0.10 & 0.566 &  472 &  28 & 0.059$\pm$0.012 \\ % 1 1
    7089 & -1.62 & -9.02 &  0.06 & 0.511 &  262 &  20 & 0.076$\pm$0.018 \\ % 1 2
\hline
\multicolumn{8}{c}{Ground-based} \\
\hline
     104 & -0.76 & -9.42 &  0.04 & 0.629 &  282 &  34 & 0.121$\pm$0.022 \\
    1261 & -1.35 & -7.81 &  0.01 & 0.593 &  320 &  47 & 0.147$\pm$0.023 \\
    2808 & -1.15 & -9.39 &  0.22 & 0.501 & 1123 &  44 & 0.039$\pm$0.006 \\
    5272 & -1.57 & -8.93 &  0.01 & 0.604 &  455 &  54 & 0.119$\pm$0.017 \\
    5904 & -1.29 & -8.81 &  0.03 & 0.573 &  325 &  33 & 0.102$\pm$0.019 \\
    6205 & -1.54 & -8.70 &  0.02 & 0.501 &  322 &  22 & 0.068$\pm$0.015 \\
    6341 & -2.28 & -8.20 &  0.02 & 0.620 &  271 &  26 & 0.096$\pm$0.020 \\
    6541 & -1.83 & -8.37 &  0.14 & 0.532 &  216 &  17 & 0.079$\pm$0.020 \\
    6779 & -1.94 & -7.38 &  0.20 & 0.585 &  259 &  16 & 0.062$\pm$0.016 \\
    7078 & -2.26 & -9.17 &  0.10 & 0.566 &  799 &  59 & 0.074$\pm$0.010 \\
\hline
\end{tabular}
\label{t:tab1}
\end{table*}

\section{Conclusions}

We derived
the relative frequency of AGB stars, given by the ratio of stars counted
on the AGB and RGB, in 26 GCs. The clusters were selected 
to have extensive and uniform high quality data and small reddening. We
found a good correlation between the ratio of AGB to RGB stars $f_{\rm AGB}=n$(AGB)/$n$(RGB)
and the minimum mass of stars along the HB $M_{\rm min}$, with a further
dependence on metallicity. This agrees with
the expectation that the less massive HB stars ($M<0.54~M_\odot$) do not even 
begin their AGB phase. For the remaining HB stars, the extension of the AGB 
lifetime is a function of their mass and metallicity. 

Since the mass of the HB stars is expected to be correlated with their
chemical composition (see Gratton et al. 2010), we then expect that most
He-rich (Na-rich and O-poor) stars in GCs do not reach the AGB. He-poor
(Na-poor, O-rich) stars may have an extended evolution on the AGB, up to
the luminosity of the tip of the RGB. The stars having an intermediate composition begin
their AGB, but terminate their AGB evolution before reaching such bright
luminosities.

As first suggested by Norris et al. (1981), this explains well the distribution 
of AGB stars along the C-N anticorrelation, which is clearly different from
that observed for RGB stars (see also Campbell et al. 2010).

\begin{acknowledgements}
%We thank the referee for her/his very interesting comments, as
%well as for reminding of the very interesting paper by Smith (1999);
%Franca D'Antona and Annibale D'Ercole for useful discussions, and 
%Valentina D'Orazi for a critical reading of the paper. 
This research has been funded by PRIN MIUR 20075TP5K9.
\end{acknowledgements}


\begin{thebibliography}{}

\bibitem[]{} Bertelli, G., Nasi, E., Girardi, L., Marigo, P. 2009, A\&A, 508, 355

\bibitem[]{} Brown, T.M., Sweigart, A.V., Lanz, T., et al. 2000, ApJ, 562, 368

\bibitem[]{} Buzzoni, A., Fusi Pecci, F., Buonanno, R., Corsi, C.E. 1983, A\&A, 128, 94

\bibitem[]{} Campbell, S.W., Lattanzio, J.C., Elliott, L.M. 2006, MSAIt, 77, 864

\bibitem[]{} Campbell, S.W., Yong, D., Wylie-de Boer, E.C. et al. 2010 in the 
proceedings of the "10th Torino Workshop on Asymptotic Giant Branch 
Nucleosynthesis: From Rutherford to Beatrice Tinsley and Beyond", Eds.: 
C. C. Worley, C. A. Tout \& R. J. Stancliffe

\bibitem[]{} Caputo, F., Castellani, V., Wood, P.R. 1978, MNRAS, 184, 377

\bibitem[]{} Carretta, E., Bragaglia, A., Gratton, R.G., et al. 2009a, A\&A, 505, 117%paper VII

\bibitem[]{} Carretta, E., Bragaglia, A., Gratton, R.G., et al. 2009b, A\&A, 508. 695

\bibitem[]{} Carretta, E., Bragaglia, A., Gratton, R.G., et al. 2010, A\&A, 516, 55

\bibitem[]{} Castellani, M., \& Castellani, V. 1993, ApJ, 407, 649

\bibitem[]{} Castellani, V., Iannicola, G., Bono, G., et al. 2006, A\&A, 446, 569

\bibitem[]{} Catelan, M. 2009, in Resolved Stellar Populations, ASP Conference
Series, D. Valls-Gabaud and M. Chavez eds, Ap\&SS, 320, 261

\bibitem[]{} D'Cruz, N.L., O'Connell, R.W., Wood, R.T., et al. 2000, ApJ, 530, 352 

\bibitem[]{} Frogel, J.A., Elias, J.H. 1988, ApJ, 324, 823

\bibitem[]{} Gratton, R., Carretta, E., Bragaglia, E., et al. 2010, A\&A, in press (arXiv 1004.3862)

\bibitem[]{} Greggio, L., Renzini, A., 1990, ApJ, 364, 35

\bibitem[]{} Harris, W.E. 1996, AJ, 112, 1487

\bibitem[]{} Lee, Y.-W., Demarque, P., Zinn, R. 1994, ApJ, 217, L101

\bibitem[]{} Mallia, E.A. 1978, A\&A, 70, 115

\bibitem[]{} Marigo, P. 2000, A\&A, 360, 617

\bibitem[]{} Moehler, S., Sweigart, A.V., Landsman, W.B., et al. 2004, A\&A, 415, 313

\bibitem[]{} Norris, J., Cottrell, P.L., Freeman, K.C., Da Costa, G.S. 1981, ApJ, 244, 205

\bibitem[]{} Paczynski, B., 1971, Acta Astron. 21, 1 

\bibitem[]{} Pilachowski, C.A., Sneden, C., Kraft, R.P., Langer, G.E. 1996, AJ, 112, 545

\bibitem[]{} Piotto, G., King, I.R., Djorgowski, S.G., et al. 2002, A\&A, 391, 945

\bibitem[]{} Reimers, D. 1975, Mem. Soc. R. Sci. Li\'ege, 6, 8, 369

\bibitem[]{} Renzini, A., Fusi Pecci, F., 1988, ARA\&A, 26, 199

\bibitem[]{} Ripepi, V., Clementini, G., Di Criscienzo, M. et al. 2007, ApJ, 667, L61 

\bibitem[]{} Rosenberg, A., Piotto, G., Saviane, I., Aparicio, A. 2000a, A\&AS, 144, 5

\bibitem[]{} Rosenberg, A., Aparicio, A., Saviane, I., Piotto, G. 2000b, A\&AS, 145, 451

\bibitem[]{} Smith, G.H., Norris, J.E. 1993, AJ, 105, 173

\bibitem[]{} Sneden, C., Ivans, I., Kraft, R.P. 2000, MSAIt, 71, 657

\bibitem[]{} Sweigart, A.V., Gross, P.G. 1976, ApJS, 32, 367

\bibitem[]{} Ventura, P. D'Antona, F., Mazzitelli, I., \& Gratton, R. 2001, ApJ, 550, L65 %IM-AGB candidati polluters

\end{thebibliography}
\end{document}